\documentclass[aip,jcp,reprint]{revtex4-1}
\usepackage{graphicx,latexsym,amsmath,amssymb}
\usepackage{color}

\def\diff{\mathop{\rm\mathstrut d\!}\nolimits}
\newcommand{\be}{\begin{equation}}
\newcommand{\ee}{\end{equation}}
\newcommand{\ba}{\begin{eqnarray}}
\newcommand{\ea}{\end{eqnarray}}

\begin{document}

\author{Gianmarco Muna\`o$^{1}$}
\thanks{Corresponding author, email: {\tt gmunao@unime.it}}
\author{Francisco G\'amez$^2$} 
\author{Dino Costa$^1$} 
\author{Carlo Caccamo$^1$}
\author{Francesco Sciortino$^3$}
\author{Achille Giacometti$^4$}
\affiliation{
$^1$Dipartimento di Fisica e di Scienze della Terra,
Universit\`a degli Studi di Messina, 
Viale F.~Stagno d'Alcontres 31, 98166 Messina, Italy \\
$^2$C/Clavel 101, Mairena del Aljarafe, 41927 Seville, Spain \\
$^3$Dipartimento di Fisica and CNR-ISC,
Universit\`a di Roma ``Sapienza'',
Piazzale Aldo Moro 2, 00185 Roma, Italy \\
$^4$Dipartimento di Scienze Molecolari e Nanosistemi,
Universit\`a Ca' Foscari Venezia, 
Calle Larga S.Marta DD2137, Venezia I-30123, Italy}

\title{Reference Interaction Site Model 
and Optimized Perturbation theories of \\
colloidal dumbbells with increasing anisotropy}

\begin{abstract}
We investigate 
thermodynamic properties
of anisotropic colloidal dumbbells in the frameworks
provided by the 
Reference Interaction Site Model (RISM) theory and an Optimized
Perturbation Theory (OPT), this latter 
based on a fourth-order high-temperature perturbative expansion 
of the free energy,
recently generalized to molecular fluids.
Our model is constituted by two identical tangent 
hard spheres surrounded by square-well attractions
with same widths and
progressively different depths.
Gas-liquid coexistence curves
are obtained by predicting 
pressures, free energies, and chemical potentials.
In comparison with previous simulation results,
RISM and OPT
agree in reproducing
the progressive reduction of the gas-liquid phase separation as
the anisotropy of the
interaction potential becomes more pronounced; in particular, the
RISM theory 
provides reasonable predictions
for all coexistence curves, bar the strong anisotropy regime,
whereas OPT performs generally less well. 
Both theories predict a linear dependence
of the critical temperature on the interaction strength, 
reproducing in this way the 
mean-field behavior observed in simulations; 
the critical density~--~that 
drastically drops as 
the anisotropy increases~---~turns to be less accurate. 
Our results appear as a robust
benchmark for further theoretical studies,
in support to the simulation approach, of self-assembly in model
colloidal systems.
\end{abstract}

\maketitle

\section{Introduction}
In recent years, a considerable number of  
experimental~\cite{Bon:14,Ma:14,Kraft:12,Nagao:11,Yoon-Chem,Forster,Hosein} 
and 
numerical~\cite{Chapela:11,Chapela2:11,Miller:09,Del-gado:11,Chong-prl}
studies have been devoted to the investigation of  
phase behavior and self-assembly
properties of colloidal dumbbells. 
Specific interest in this class
stems essentially from the possibility to accurately tune the 
aspect ratio of the constituting spheres, as well as their interaction
properties, so to obtain rich and fascinating phase 
behaviors.\cite{Munao:14,Munao:15,Avvisati-JCP} In particular, if one of the
two particles is solvophilic and the other one is solvophobic, 
colloidal dumbbells
represent a simple example of surfactant (Janus 
dumbbells),\cite{Janusdumbbell,Janusdumbbell1} constituting 
a molecular generalization of Janus spherical 
particles,\cite{granick,Janusweitz,Adv:10} largely investigated in the
last years because of their peculiar self-assembly properties and phase
behaviors.\cite{cacciuto,Chen:11,Granick-Lang,%
Granick-Nat,janusprl,januslong,Vissers:jcp}

While a large number of experimental and numerical investigations 
have involved colloidal dumbbells, 
considerable less attention has been paid to theoretical studies:
the framework of potential energy landscape
has been adopted to describe the self-assembly of dumbbells into
helices;\cite{Fejer:14} 
the competition between self-assembly and phase separation
has been documented by integral equation
theories;\cite{Munao:PCCP} the same approach~\cite{Munao-cpl,Munao:09}
and fundamental measure theory~\cite{Marechal:11}
have been adopted to investigate
structure and thermodynamics of hard dumbbells.
Notwithstanding all such studies, a systematic theoretical 
description of the phase behavior 
of colloidal dumbbells is still lacking. 
In order to fill this gap, here we 
calculate systematically 
free energy, pressure, chemical potential and phase diagram
of anisotropic colloidal dumbbells,  so
to investigate how such properties change by 
tuning the interaction potential.
We consider to this task a series
of model dumbbells constituted by two identical 
tangent hard spheres 
surrounded by square-well attractions 
with same widths and progressively different depths.
We carry out our investigation 
in the frameworks provided by
the Reference
Interaction Site Model theory (RISM)\cite{chandler:1972} 
and a recently proposed Optimized 
Perturbation Theory~(OPT).\cite{DelRio:09,Gamez:14} 
Theoretical predictions for the 
gas-liquid phase coexistence are compared
with previous Monte Carlo results by us\cite{Munao:14}
and other authors.\cite{Cai:12} The present work extends
our preliminary OPT analysis of
the symmetric case~\cite{Gamez:14} and
our RISM study on a slightly different
dumbbell model.\cite{Munao:PCCP}

As for the 
RISM theory,\cite{chandler:1972} it constitutes
a molecular generalization of the 
Ornstein-Zernike theory of simple fluids.\cite{Hansennew}
Originally devised 
for rigid molecules constituted by hard spheres,\cite{lowden:5228}
the theory was later 
extended to more complex systems, like colloidal models;
in particular, it was used to investigate
the thermodynamic and structural properties of
discotic lamellar colloids,\cite{Harnau:01,Harnau:05}
the self-assembly in diblock copolymers modeled
as ``ultrasoft'' colloids,\cite{Hansen:06}
the interaction between colloidal particles and
macromolecules,\cite{Khalatur:97}
the crystallization and solvation properties
of nanoparticles in aqueous solutions,\cite{Kung:10} the
liquid structure of tetrahedral colloidal particles\cite{Munao:11},
and the self-assembly properties of Janus rods.\cite{Tripathy-RISM}

As for 
the perturbation theory, this scheme played a significant
role in describing the thermodynamic properties of a
large variety of 
fluids.\cite{Valadez:12,Gamez:11,Benavides:07,Elliott:05,Cui:02,Chapela:89} 
The theory
was originally proposed as a high-temperature expansion
by several groups and then recast in a very efficient computational tool
by Barker and Henderson.\cite{BH1,BH2,BH3} It has been recently adapted to
cope with Janus~\cite{Giacometti:14} and tri-block Janus~\cite{Gogelein:12} 
fluids.
Here we use an optimized version of this theory
based on a
fourth-order high-temperature perturbative expansion
of the free energy, recently developed to 
fit a large set of simulation data for square-well fluids.\cite{DelRio:09}
Recently, one of the authors generalized 
this approach to deal with more complex systems, like
molecular fluids.\cite{Gamez:14} 
Within this formalism, the effective energy
and range of square-well interactions 
depend on the molecular anisotropy.
In this way,
it is possible to calculate analytical equations of state 
for a large variety of molecular models
like for instance dimers, trimers, chains and spherocylinders.\cite{Gamez:14}

The importance of assessing the accuracy of these theoretical
approaches is crucial, as a large number of 
demanding numerical calculations
\cite{Munao:15} is typically required to investigate the 
experimentally relevant heterogeneous
dumbbells case.\cite{Kraft:12} A reliable theoretical tool 
would then help to focus on those systems of great interest.

The paper is organized as follows: in Section II we provide the details of 
interaction potential and RISM and OPT schemes.
Results are
presented and discussed in Section III and conclusions follow in Section IV.

\begin{figure}
\begin{center}
\begin{tabular}{cc}
\includegraphics[width=4.0cm,angle=-180]{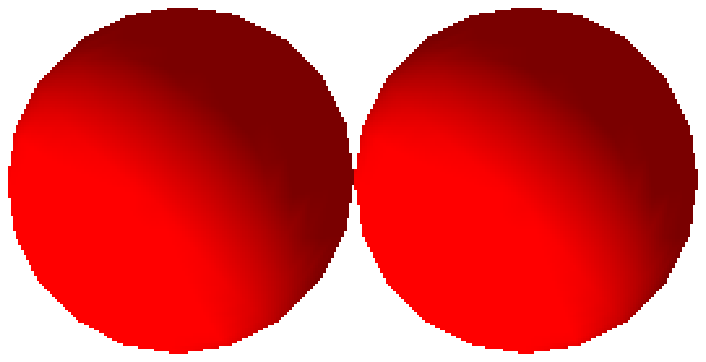}
\includegraphics[width=4.0cm,angle=-180]{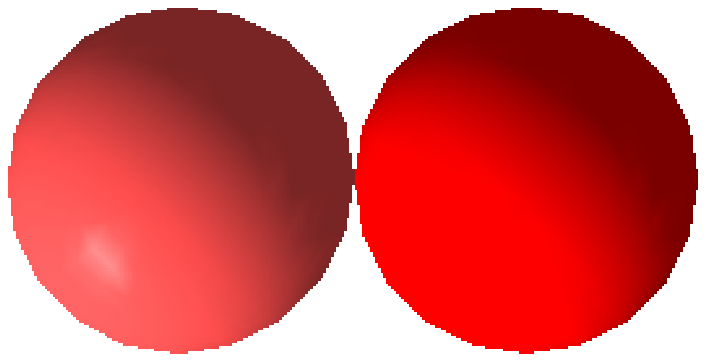} \\
\includegraphics[width=4.0cm,angle=-180]{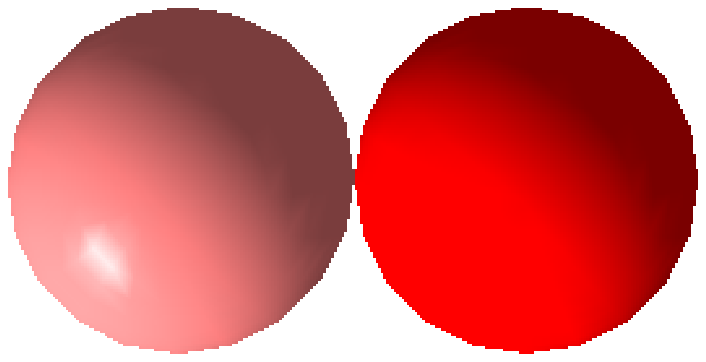} 
\includegraphics[width=4.0cm,angle=-180]{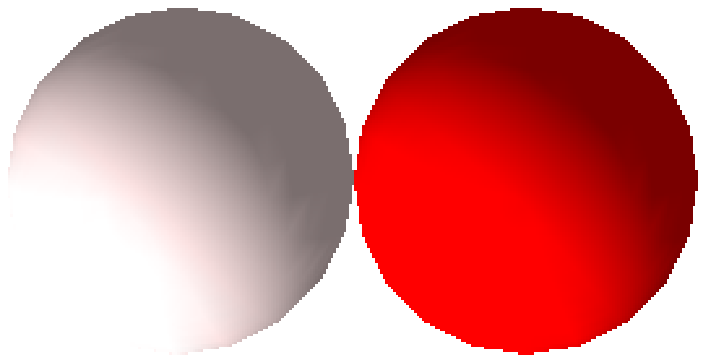}
\end{tabular}
\caption{Schematic representation of colloidal dumbbells investigated in this
work. Starting 
from the symmetric case 
$\varepsilon_{11}=\varepsilon_{12}=\varepsilon_{22}$ (top, left), 
the interaction becomes progressively asymmetric
by reducing 
$\varepsilon_{11}=\varepsilon_{12}=0.7$, 0.5, 0.3, $0.1\varepsilon_{22}$.
The weakening of $\varepsilon_{11}$ 
is signaled
by progressively ``fading'' the color of the first site.}\label{fig:dumb}
\end{center}
\end{figure}

\section{Models and theories}
\subsection{Interaction potentials}
The sequence of models investigated in this work
is schematically represented  in Fig.~\ref{fig:dumb}: they are constituted
by two identical tangent hard
spheres of diameter $\sigma$, labeled as 1 and 2,
interacting with sites 1 and 2 of 
another dumbbell at distance $r$
via square-well (SW) attractions with same width ($\lambda=0.5$)
and different depths $\varepsilon_{ij}$ $(i,j=1,2)$:
\ba\label{eq:vij}
V_{ij}(r)=\begin{cases}
\infty      & \mbox{if } r < \sigma       \\[4pt]
-\varepsilon_{ij} & \mbox{if } \sigma \le r < \sigma+\lambda\sigma \\[4pt]
0           & \mbox{otherwise}\,.
\end{cases}
\ea
Parameters
$\sigma$ and $\varepsilon_{22}$
provide, respectively,
the unit of length
and energy, in which terms
we define the reduced temperature $T^*=k_{\rm B}T/\varepsilon_{22}$
(with $k_{\rm B}$ as the Boltzmann constant), number
density $\rho^*=\rho\sigma^3$ and pressure $P^*=P\sigma^3/\varepsilon_{22}$. 
In all calculations we have fixed
$\varepsilon_{12} (\equiv\varepsilon_{21})=\varepsilon_{11}$;
then, 
we have studied
a sequence of cases 
obtained by progressively reducing $\varepsilon_{11}$
starting from the symmetric case,
$\varepsilon_{11}= \varepsilon_{22}$,
to $\varepsilon_{11}=0.7$, 0.3, 0.1 (in units of $\varepsilon_{22}$).

\subsection{Reference interaction site model theory}
In the RISM framework\cite{chandler:1972}
the pair structure of a fluid composed by
identical two-site molecules
is characterized
by a set of four site-site intermolecular
pair correlation functions $h_{ij}(r)=g_{ij}(r)-1$
where $(i,j)=(1,2)$ and
$g_{ij}(r)$ are the site-site radial distribution functions.
The $h_{ij}(r)$
are related to a set of intermolecular direct correlation
functions $c_{ij}(r)$  by a matrix generalization of the 
Ornstein-Zernike equation for simple fluids,~\cite{Hansennew} expressed 
in the $k$-space as:
\ba\label{eq:rism}
{\bf H}(k) = {\bf W}(k){\bf C}(k){\bf W}(k) +
\rho{\bf W}(k){\bf C}(k){\bf H}(k)\,, 
\ea
where
${\mathbf H}\equiv [h_{ij}(k)]$,
${\mathbf C}\equiv[c_{ij}(k)]$, and
${\mathbf W}\equiv [w_{ij}(k)]$
are $ 2 \times 2$ symmetric matrices;
the elements $w_{ij}(k)$
are the Fourier transforms
of the intramolecular correlation functions, written explicitly as:
\begin{eqnarray}\label{eq:intra}
w_{ij}(k) =
\frac{\sin[kL_{ij}]}{kL_{ij}}\,,
\end{eqnarray}
where the bond length $L_{ij}$ is given either
by  $L_{ij}=\sigma$, if $i\ne j$, or by $L_{ij}=0$, otherwise.
In order to solve the RISM equation~(\ref{eq:rism}),
we have adopted in this work the Kovalenko-Hirata (KH)
closure~\cite{Kovalenko:99,Kovalenko:01} that assumes~---~
beside the exact expression $g_{ij}(r)=0 \mbox{ if } r \leq \sigma$~---~%
the direct correlation functions outside the hard-core
to be approximated by a combination
of the Mean Spherical Approximation (MSA)
and HyperNetted Chain (HNC) expression:\cite{Hansennew}
\ba\label{eq:kh}
 c_{ij}(r) & = & 
\begin{cases}
\mbox{HNC} & \mbox{if } g_{ij}(r) \leq 1 \\
\mbox{MSA} & \mbox{if } g_{ij}(r)  >   1 
\end{cases} \nonumber\\
&\qquad \equiv & 
\begin{cases}
\exp[- \beta V_{ij}(r) + \gamma_{ij}(r)] - \gamma_{ij}(r) - 1\\
-\beta V_{ij}(r)\,,
\end{cases}
\ea
where $\beta=1/T^*$ and $\gamma_{ij}(r)=h_{ij}(r) - c_{ij}(r)$.
We have implemented the numerical solution of the RISM/KH
scheme by means of  a standard iterative Picard algorithm,
on a grid of 8192 points with a mesh $\Delta r=0.005\sigma$. 

In order to 
calculate the thermodynamic properties, 
we have used standard thermodynamic 
integration formul{\ae}~\cite{Hansennew} that 
already proven to be successful for
SW dumbbells.\cite{Munao:PCCP} Specifically,
we have calculated the excess free energy per particle via
the energy route
along constant-density paths according to:
\begin{eqnarray}\label{eq:ba}
\frac{\beta F^{\rm ex}(\beta)}{N} & = &  \frac{\beta F^{\rm ex}(\beta=0)}{N}
+ \int_{0}^{\beta} \frac{U(\beta')}{N}
{\diff \beta'}\,,
\end{eqnarray}
where $U(\beta)$ is the internal energy of the fluid
at a given inverse temperature $\beta$ and $F^{\rm ex}(\beta=0)$
corresponds to the excess free energy of tangent hard-sphere dumbbells.
For this, the most classical expressions
known in the literature are an
analytic expression due to
Tildesley and Street~\cite{Tildesley:80} (and devised to
fit their Monte Carlo data) and the Wertheim equation of state 
for tangent hard dumbbells,\cite{WerthPolym} reading:
\begin{eqnarray}\label{eq:Werth}
\frac{\beta F^{\rm ex}(\beta=0)}{N}&=&n\frac{(4\eta-3\eta^2)}{(1-\eta)^2}  \nonumber\\[4pt]
&+&   (n-1)\left [ 3\ln(1-\eta)+\ln\left ( \frac{2}{2-\eta} \right ) \right ] \nonumber\\[4pt]
&+&\ln\left (  {6\eta/\pi}\right )-1\,,
\end{eqnarray}
where $\eta=2(\pi/6)\rho^*$ is the dumbbell packing fraction.
We have explicitly verified that such equation agrees with the free energy
expression for hard dumbbells provided in Ref.~\onlinecite{Tildesley:80}.
Equation~(\ref{eq:Werth}) has the additional advantage to avoid free 
parameters and the 
fit of simulation data, so that we will use it through all the calculations.

The excess pressure is then obtained through the (numerical)
derivative of the l.h.s. of Eq.~(\ref{eq:ba}) with respect to the density,
taken at constant temperature, while coexistence conditions are determined
by searching for equal chemical potentials and pressures 
(at a given temperature)
of the two coexisting phases.
Finally,
critical temperatures and densities are obtained by fitting
the coexistence points through the scaling law for the
densities and the law of rectilinear diameters with an effective critical
exponent $\delta$=0.32.\cite{frenkelsmit}
 We refer the
reader to our previous work~\cite{Munao:PCCP}
for full details on the above procedure.

Using the symmetric model as a test case, we have also 
investigated the quality of predictions obtained by 
different approaches within the RISM framework. Specifically, we have
studied the HNC closure, that assumes
expressions in the top lines of Eq.~(\ref{eq:kh}) holding for all $r> \sigma$
independently on the value of $g(r)$. Moreover, within the KH
closure, we have studied a different route from structure to thermodynamics,
based on closed formul{\ae} (i.e requiring no thermodynamic integration)
to calculate free energy and pressure.
Such expressions, derived by Kovalenko and Hirata in
Refs.~\onlinecite{Kovalenko:99,Kovalenko:01},
closely follow the corresponding
relations deduced long ago~\cite{Morita:60} within the Ornstein-Zernike/HNC 
scheme for simple fluids and later generalized 
to the RISM/HNC scheme.\cite{Singer:85} 
We refer the 
reader to the 
original works and our recent application~\cite{Munao:11}
for full details.

\subsection{Optimized perturbation theory}

The starting point of OPT is given by
the optimized equation of state 
recently proposed in Ref.~\onlinecite{DelRio:09}, where
the free energy of a fluid composed by SW particles
with well depth $\varepsilon$ and 
width $\lambda$,
at temperature $T$ and density $\rho$ is expressed as a 
perturbation series in the form:
\be\label{eq:sw}
\frac{\beta F(\beta)}{N} = \frac{\beta F(\beta=0) }{N}
+ \sum_{m=1}^\infty 
\left[\frac{\varepsilon}{k_{\rm B}T}\right]^m f_{m}
(\rho ,\lambda)\,.
\ee
In Eq.~(\ref{eq:sw}) $F^{\rm ex}(\beta=0) $
is the free energy of the reference hard-sphere fluid
plus the ideal contribution,
and $f_m$  are $m$-th order perturbation terms. 
In Ref.~\onlinecite{DelRio:09}
the series is truncated, and $f_m$ are calculated up
to the fourth-order.
The theory
includes 
the Barker-Henderson~\cite{BH1} result as a truncation
of the expansion to second order. The third and fourth orders, 
however, are not included as explicit higher order corrections 
to the Barker-Henderson theory, that would require uncontrolled 
truncations of the hierarchy of correlation functions, 
but as an optimized phenomenological
calculation of coefficients,
hinging on simulations results.~\cite{DelRio:09} 
It turns out that this theory for the  SW fluid is 
accurate
over nearly the whole density range and subcritical (low) temperatures.
We refer the reader to the original work~\cite{DelRio:09} for
detailed expressions of $f_m$ $(m=1,\ldots,4)$, as well as all other
necessary machinery.

The OPT generalization at issue,
pertinent to a fluid
composed by SW dumbbells, is written explicitly as
a sum of SW contributions, each taking separately into account
the four $(i,j=1,2)$
interactions of Eq.~(\ref{eq:vij}), in the form:\cite{Gamez:14}
\be\label{eq:adis}
\frac{\beta F(\beta)}{N} =\frac{\beta F(0) }{N}+
\sum_{i,j=1}^2 \left\{
\sum_{m=1}^4\,
\left[\frac{\varepsilon_{ij}}{k_{\rm B}T}\right]^m f_{m}
(\rho ,\lambda)\right\}\,.
\ee
As visible, each term in the free energy expansion
is weighted by a temperature reduced
by the corresponding well depth $\varepsilon_{ij}$, 
$\rho$ is the density of sites 1 or 2, and $\lambda$ takes 
the fixed value 0.5.
Given the spherical symmetry of each site-site interaction, 
the molecular anisotropy is taken into account by the reference free energy,
provided in our case by the Wertheim 
equation of state for tangent hard
dumbbells of Eq.~(\ref{eq:Werth}).\cite{WerthPolym}.
This choice gives better results for 
non-convex particles in comparison with the possibility to define
an effective hard-sphere diameter with the corresponding 
scaling of density.\cite{Gamez:14};
it turns to be accurate for a moderate number of beads per chain.

Given Eq.~(\ref{eq:adis}),
one can obtain the thermodynamic properties from usual 
thermodynamic relations. For instance, the compressibility factor
\begin{eqnarray}
Z\equiv \frac{\beta P}{\rho}= 1 + \eta \frac{\partial}{\partial \eta} 
\left[  \frac{\beta F^{\rm ex}}{N}\right]_T
\end{eqnarray}
can also be expressed as a  high-temperature expansion,
\be
\label{zeta}
  Z =1+Z^{\rm ex}(\beta=0) +
\sum_{i,j=1}^2 \left\{
\sum_{m=1}^{\infty } 
\left[\frac{\varepsilon_{ij}}{k_{\rm B}T}\right]^m z_{m}
(\rho ,\lambda)\right\}\,,
\ee
where $z_m=\eta \partial/\partial \eta \left[  
f_m(\rho,\lambda)\right]_T$.

As in the RISM framework, we have determined
the gas-liquid phase boundaries 
by the condition of thermal, mechanical, and chemical equilibrium.
Numerical methods have been used to this task,
based in this case on a combination of the
Levenberg-Marquardt algorithm with Gauss and steepest-descent
methods.\cite{Meeter}

\section{Results}

\begin{figure}
\begin{center}
\includegraphics[width=9.0cm,angle=0]{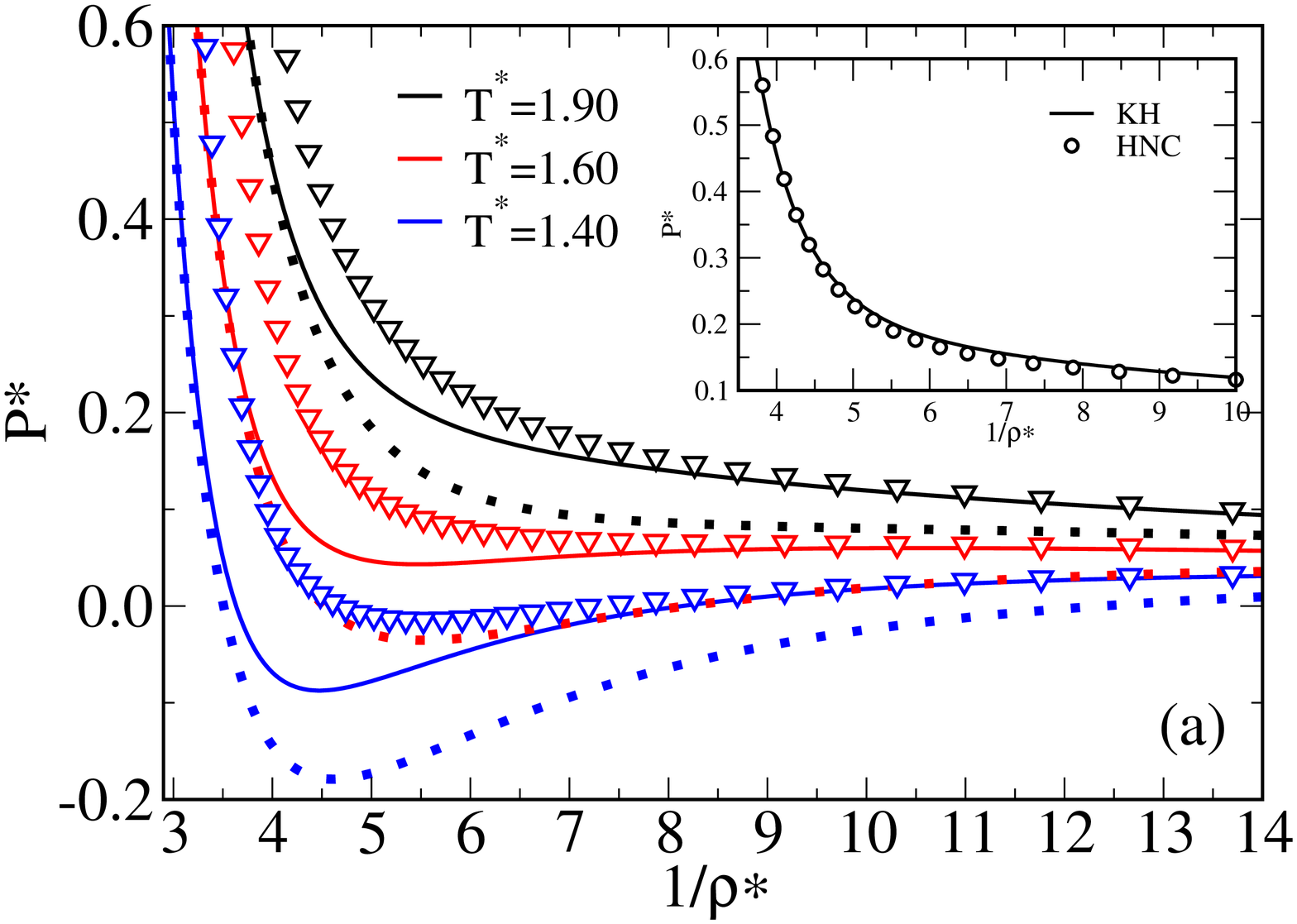}  \\
\includegraphics[width=9.0cm,angle=0]{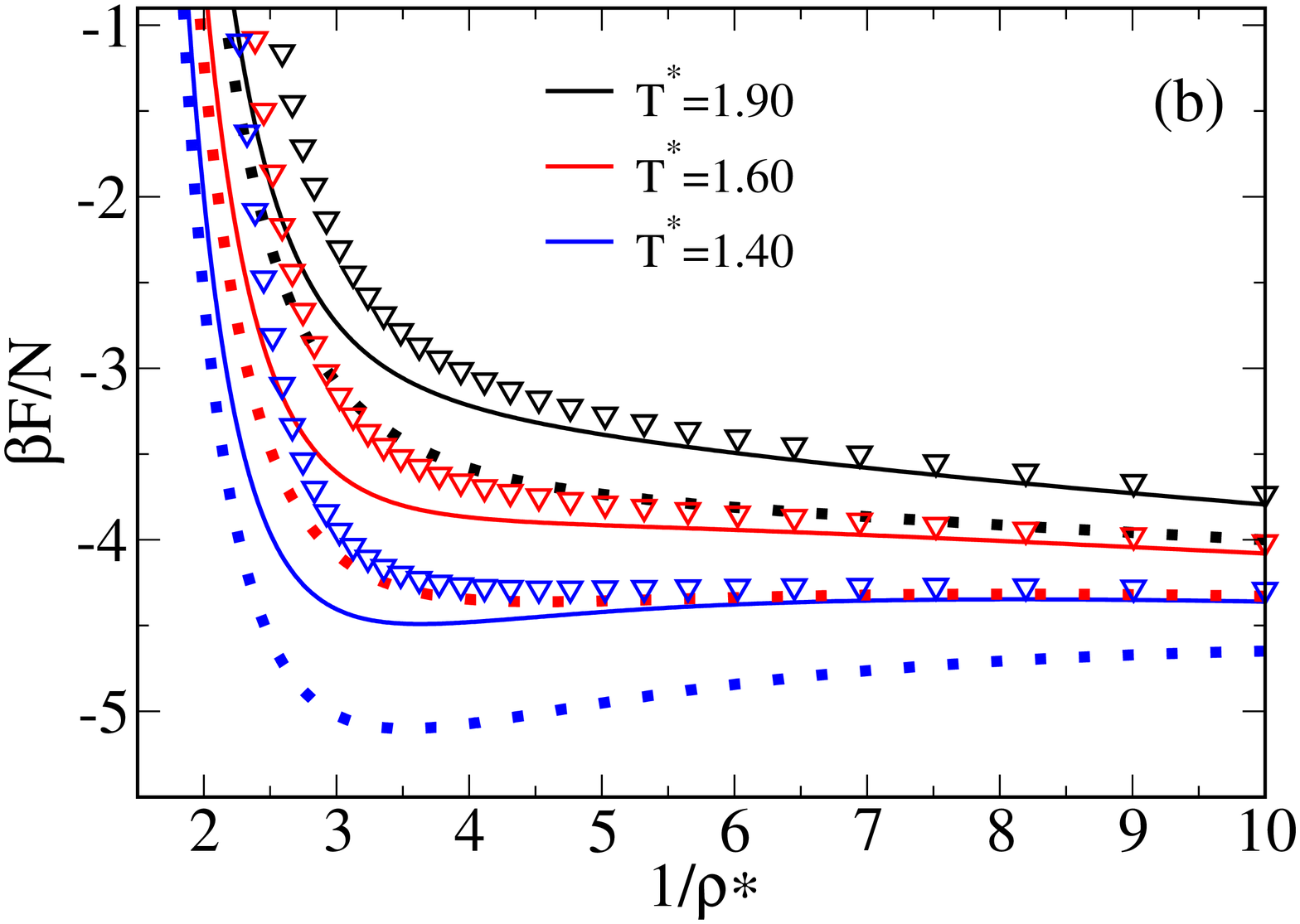} \\ 
\includegraphics[width=9.0cm,angle=0]{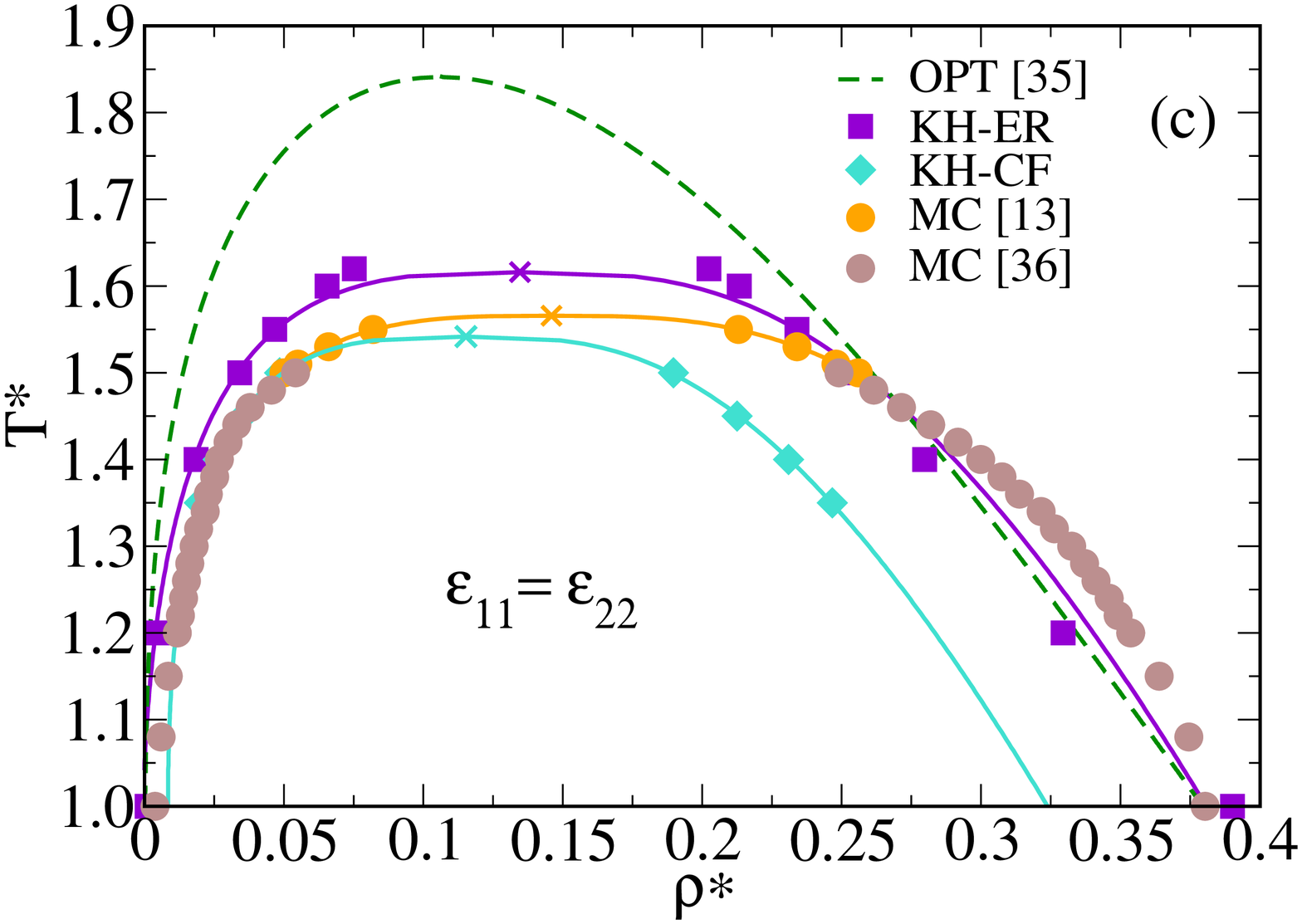}
\caption{Pressure (a) and free energy (b)
of symmetric colloidal dumbbells along different isotherms (see legends): 
KH energy route (lines), KH closed formul{\ae}
(triangles) and OPT (small squares).
In the inset, KH and HNC 
energy-route pressures at $T^*=1.90$.
Panel c: gas-liquid coexistence by
KH energy-route (KH-ER, squares), KH closed formul{\ae} (KH-CF, diamonds),
OPT (dashed line)~\cite{Gamez:14} and
MC~(circles).\cite{Cai:12,Munao:14} 
Full lines are best-fits of KH and simulation points (see text)
with corresponding critical points (crosses).}\label{fig:epsi1}
\end{center}
\end{figure}
%
In Fig.~\ref{fig:epsi1} we compare the energy-route KH and OPT predictions
for symmetric colloidal dumbbells,
i.e. with $\varepsilon_{11}=\varepsilon_{22}$
in Eq.~(\ref{eq:vij}).
Results for the pressure (a) and free energy (b)
are reported along different isotherms across the critical
temperature $T^*_{\rm c} \approx 1.60$
(estimated in our previous work).\cite{Munao:14}
KH and OPT predict the same trends
both for the pressure and free energy;
in particular, they provide evidence
for a supercritical fluid at $T^*=1.90$, whereas a van der Waals 
loop~---~heralding a gas-liquid phase separation~---~develops
at $T^*=1.60$ and
becomes well defined at $T^*=1.40$.
The quantitative agreement between the two theories
slightly worsens
at low temperatures, where
OPT curves fall progressively below
the corresponding KH ones;
this circumstance implies a higher OPT critical temperature and,
correspondingly, a wider gas-liquid coexistence curve. 
In Fig.~\ref{fig:epsi1}c we compare
KH gas-liquid coexistence curves~---~as obtained by the energy route and the
closed formul{\ae}~---~with Monte Carlo (MC) data 
by us~\cite{Munao:14} and other authors.\cite{Cai:12} 
OPT predictions, already provided in Ref.~\onlinecite{Gamez:14}, 
are also reported for 
comparison. One can notice that KH reproduces reasonably well
all features, whereas OPT, as observed, overestimates
the two-phase region.
Such a discrepancy 
could be possibly ascribed to the truncation involved in the OPT procedure
(see Eq.~(\ref{eq:adis})). In this case improvements could be achieved
by incorporating higher order terms in the free energy expansion,
so to take into account more accurately density fluctuations,
along the lines described for instance in Ref.~\onlinecite{Salvino:92}.
In general, the OPT scheme for SW dumbbells
shares the same level of accuracy
of the original approach for the SW atomic fluid,~\cite{DelRio:09}
the latter performing slightly better in the critical region.

\begin{figure}[!t]
\begin{center}
\includegraphics[width=9.0cm,angle=0]{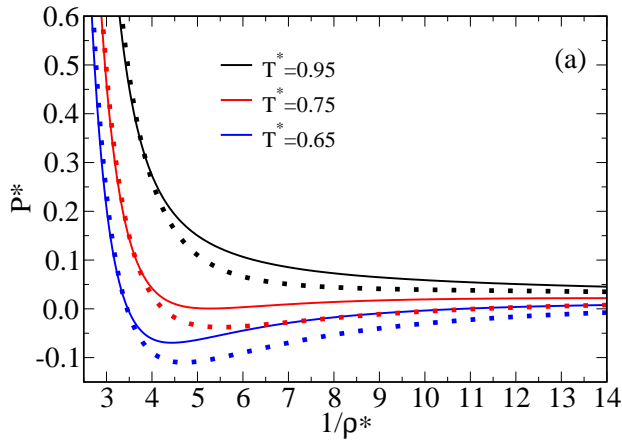}  \\
\caption{KH (lines) and OPT (small squares) pressures 
of colloidal dumbbells with $\varepsilon_{11}=0.3\varepsilon_{22}$.
}\label{fig:epsi03} 
\end{center}
\end{figure}

As for HNC, we have verified that 
at high temperatures 
no significant differences arise with the pressure calculated
within the KH scheme (see the curves corresponding to $T^*=1.90$
in the inset of Fig.~\ref{fig:epsi1}a).
On the other hand, upon lowering the temperature,
the numerical convergence of 
HNC becomes progressively more problematic, thus 
precluding from further exploration 
a progressively larger interval of
$T-\rho$ conditions.
As a consequence, the conditions for gas-liquid coexistence 
turn to be completely unattainable.
This issue is even more
pronounced as the molecular anisotropy increases,
leaving KH and OPT schemes as the only reliable
approaches to study the phase separation properties of our models. 
\begin{figure}[!t]
\begin{center}
\begin{tabular}{c}
\includegraphics[width=9.0cm,angle=0]{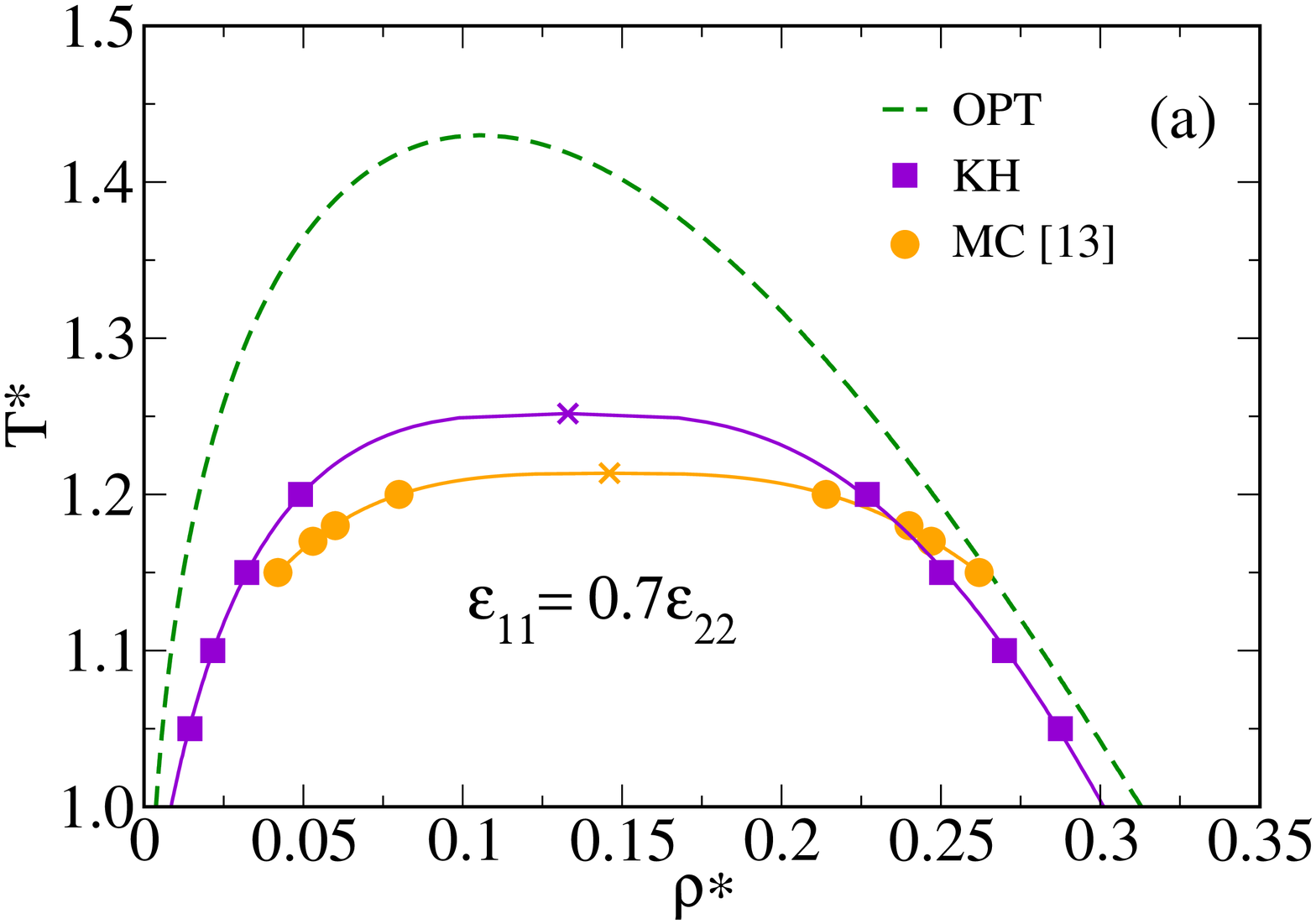} \\
\includegraphics[width=9.0cm,angle=0]{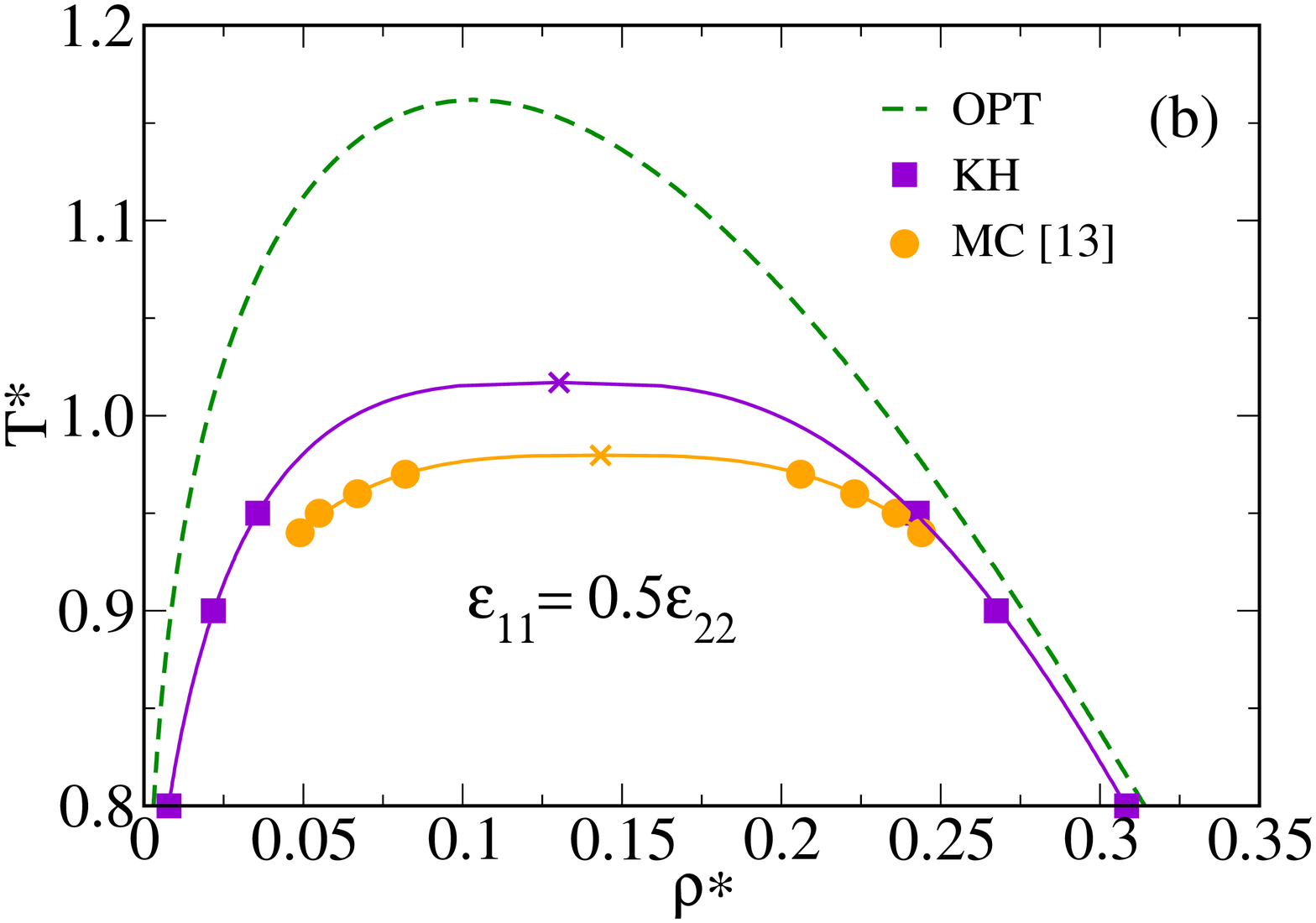} \\
\includegraphics[width=9.0cm,angle=0]{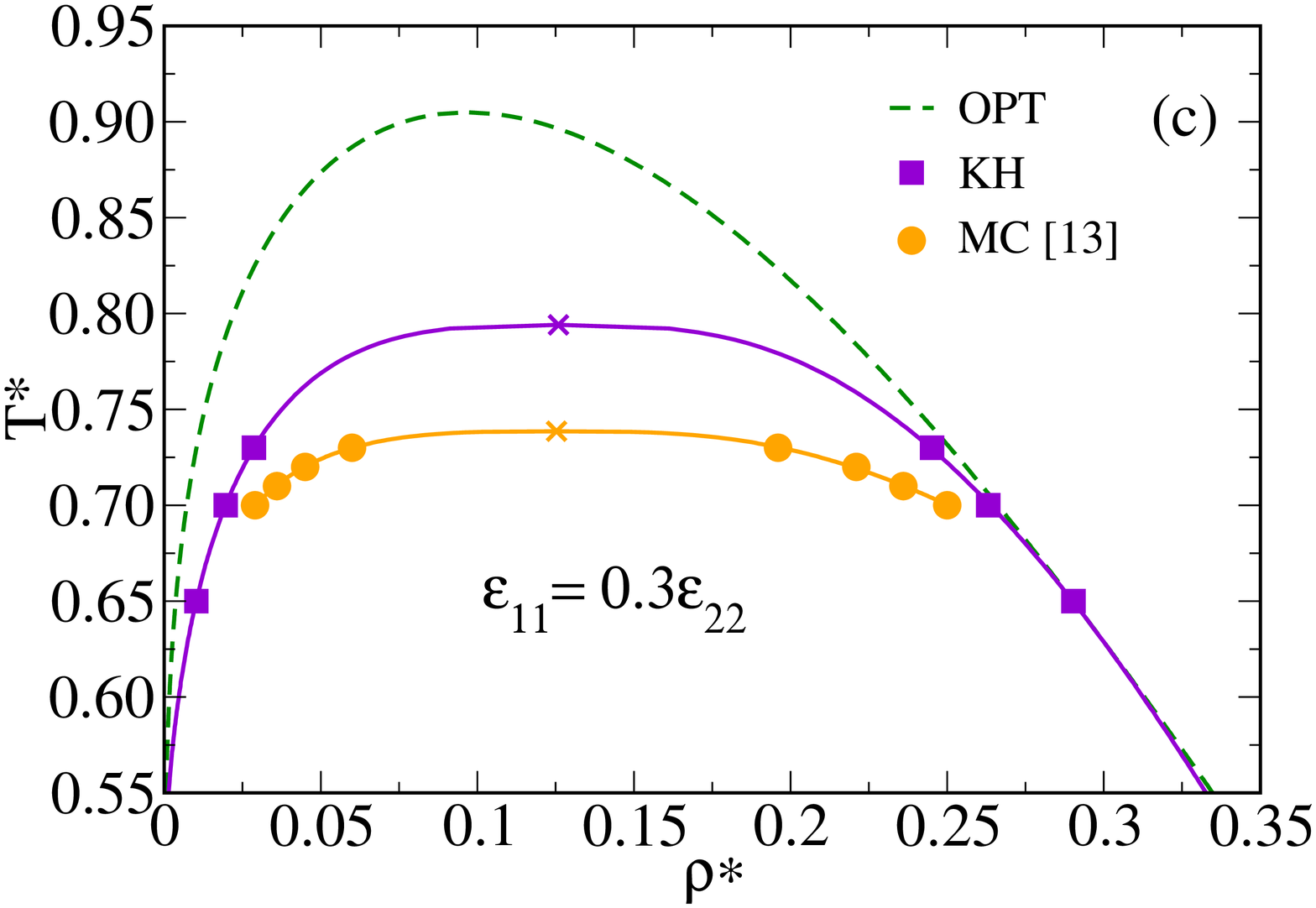}
\end{tabular}
\caption{Gas-liquid coexistence curves
of colloidal dumbbells upon progressively decreasing $\varepsilon_{11}$.
}\label{fig:binodal} 
\end{center}
\end{figure}

As for KH closed formul{\ae},
pressures and free energies at different temperatures
are also reported in Fig.~\ref{fig:epsi1}.
As visible, a van der Waals loop in the pressure
is observed in the same range
of temperatures of KH energy route and OPT, even if
systematically higher and shallower estimates 
are obtained. 
The same discrepancy appears to affect the predictions for the free energy.
As a result (see Fig.~\ref{fig:epsi1}c), 
the liquid branch of phase coexistence
appears too low in comparison with simulations, KH 
energy route and OPT, whereas
the critical temperature and the whole gas branch turn to be
 quite well predicted.
Overall, the KH energy route 
appears more accurate and therefore we shall adopt such a 
scheme%
~---~together with the differently based OPT approach~---~%
to investigate the phase behavior of other
cases with $\varepsilon_{11}<\varepsilon_{22}$.

As visible from Fig.~\ref{fig:epsi03},
the agreement between KH and OPT predictions
for the pressure essentially persists as 
$\varepsilon_{11}$
is reduced down to $0.3\varepsilon_{22}$.
In Ref.~\onlinecite{Munao:14}
we have documented 
the progressive shrinkage of the coexistence curves toward 
low temperatures as 
$\varepsilon_{11}$ is reduced. 
As visible from Fig.~\ref{fig:binodal},
where coexistence curves for $\varepsilon_{11}=0.7$, 0.5, and 
$0.3\varepsilon_{22}$ are reported, 
this feature is reproduced by both theories,
essentially at the same level of accuracy observed for 
the symmetric case: 
KH still provides reasonable estimates of the
gas-liquid coexistence curve, with critical temperatures and
densities fairly well predicted, whereas
OPT overestimates the amplitude of the
two-phase region
and, accordingly, the critical temperature. 

\begin{figure}
\begin{center}
\includegraphics[width=9.0cm,angle=0]{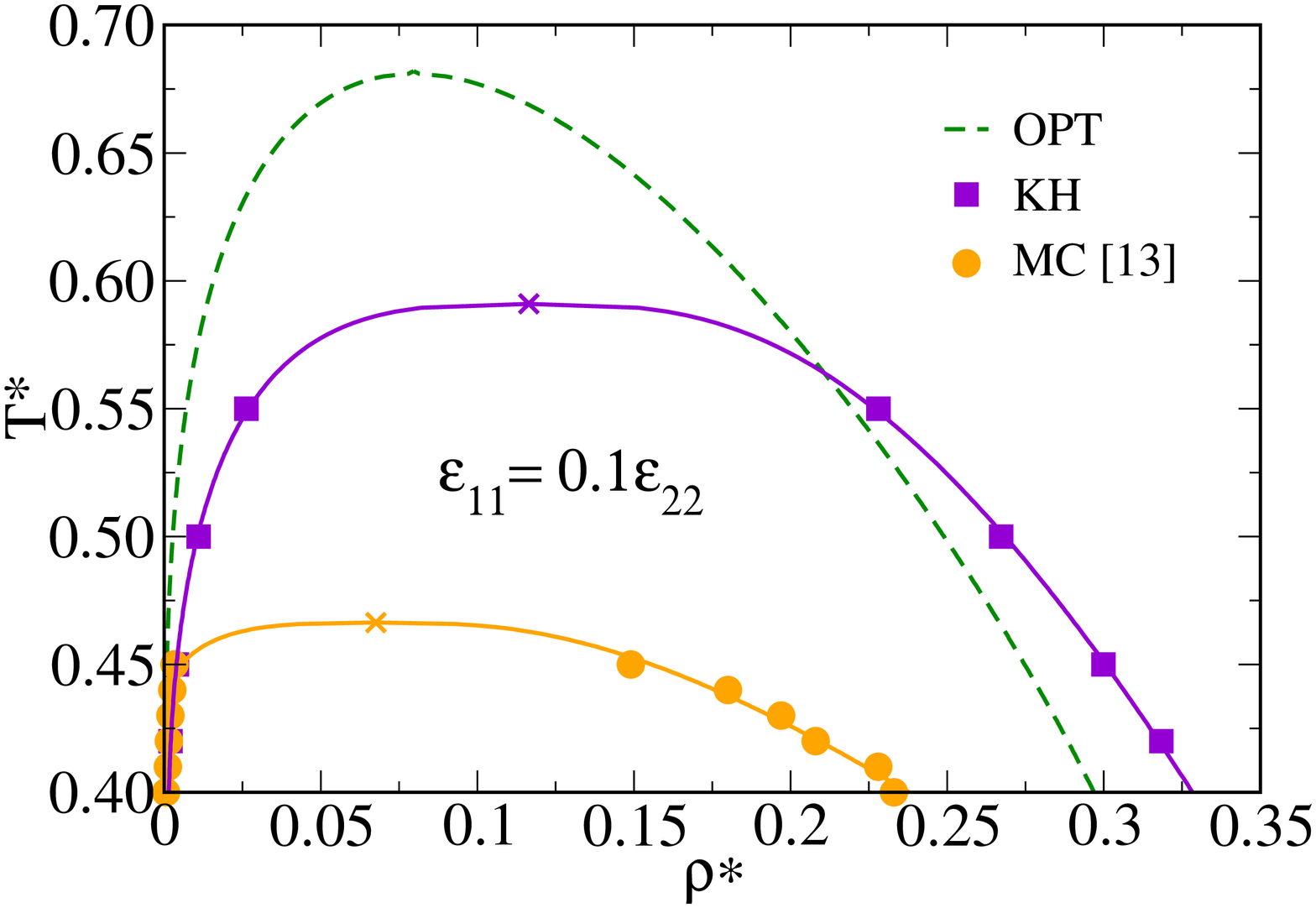}
\caption{Gas-liquid coexistence of colloidal dumbbells with 
$\varepsilon_{11}=0.1\varepsilon_{22}$.
}\label{fig:epsi01} 
\end{center}
\end{figure}

As documented in our recent simulation study,\cite{Munao:14}
increasing the asymmetry of total interaction
promotes the development of aggregates in the fluid; specifically
at $\varepsilon_{11}=0.1\varepsilon_{22}$, 
dumbbells self-assemble into micelles
at low densities and planar structures (lamell\ae) at intermediate/high
densities, provided the temperature is low enough.
Moreover 
the self-assembly
process tends to destabilize 
the gas-liquid phase separation~--~completely disappearing 
for Janus dumbbells, i.e. with
$\varepsilon_{11}=0$~---~with 
the coexistence curve stretched toward very low 
gas densities and
a critical temperature drastically lower 
than in the symmetric case. 
At variance with previous cases, both KH and OPT theories
fail to reproduce such a behavior, largely
overestimating the coexistence region
at $\varepsilon_{11}=0.1\varepsilon_{22}$, as visible in Fig.~\ref{fig:epsi01}.

Indeed, we may surmise 
that liquid state theories, explicitly tailored to
describe a homogeneous fluid environment,
progressively worsen as local 
inhomogeneities tend to develop in the system.
Also, the progressive displacement of binodal curves
 toward lower temperatures may negatively 
affect the OPT predictions based, as described, on 
a  high-temperature expansion of the free energy.

\begin{table}[!t]
\caption{\label{tab:crit}
Critical temperatures, densities and chemical potentials 
for colloidal dumbbells with variable $\varepsilon_{11}$
as obtained by MC,~\cite{Munao:14} KH and OPT.
}
\begin{center}
\begin{tabular*}{0.45\textwidth}{@{\extracolsep{\fill}}clccc}
\hline
$\varepsilon_{11}/\varepsilon_{22}$ & & $T^*_{\rm c}$ & $\rho^*_{\rm c}$ & 
$\mu_{\rm c}/\varepsilon_{22}$ \\
\hline
1.0 & MC  & 1.566 & 0.146 & -5.680 \\
    & KH  & 1.616 & 0.135 & -5.973 \\
    & OPT & 1.842 & 0.106 & -6.549 \\[4pt]
0.7 & MC  & 1.213 & 0.146 & -4.524 \\
    & KH  & 1.252 & 0.133 & -4.610 \\
    & OPT & 1.430 & 0.105 & -5.025 \\[4pt]
0.5 & MC  & 0.980 & 0.143 & -3.725 \\
    & KH  & 1.017 & 0.130 & -3.762 \\
    & OPT & 1.162 & 0.103 & -4.107 \\[4pt]
0.3 & MC  & 0.739 & 0.125 & -2.997 \\
    & KH  & 0.794 & 0.126 & -2.964 \\
    & OPT & 0.905 & 0.097 & -3.248 \\[4pt]
0.1 & MC  & 0.466 & 0.068 & -2.635 \\
    & KH  & 0.591 & 0.116 & -2.284 \\
    & OPT & 0.682 & 0.080 & -2.205 \\
\hline
\end{tabular*}
\end{center}
\end{table}

\begin{figure}
\begin{center}
\begin{tabular}{c}
\includegraphics[width=9.0cm,angle=0]{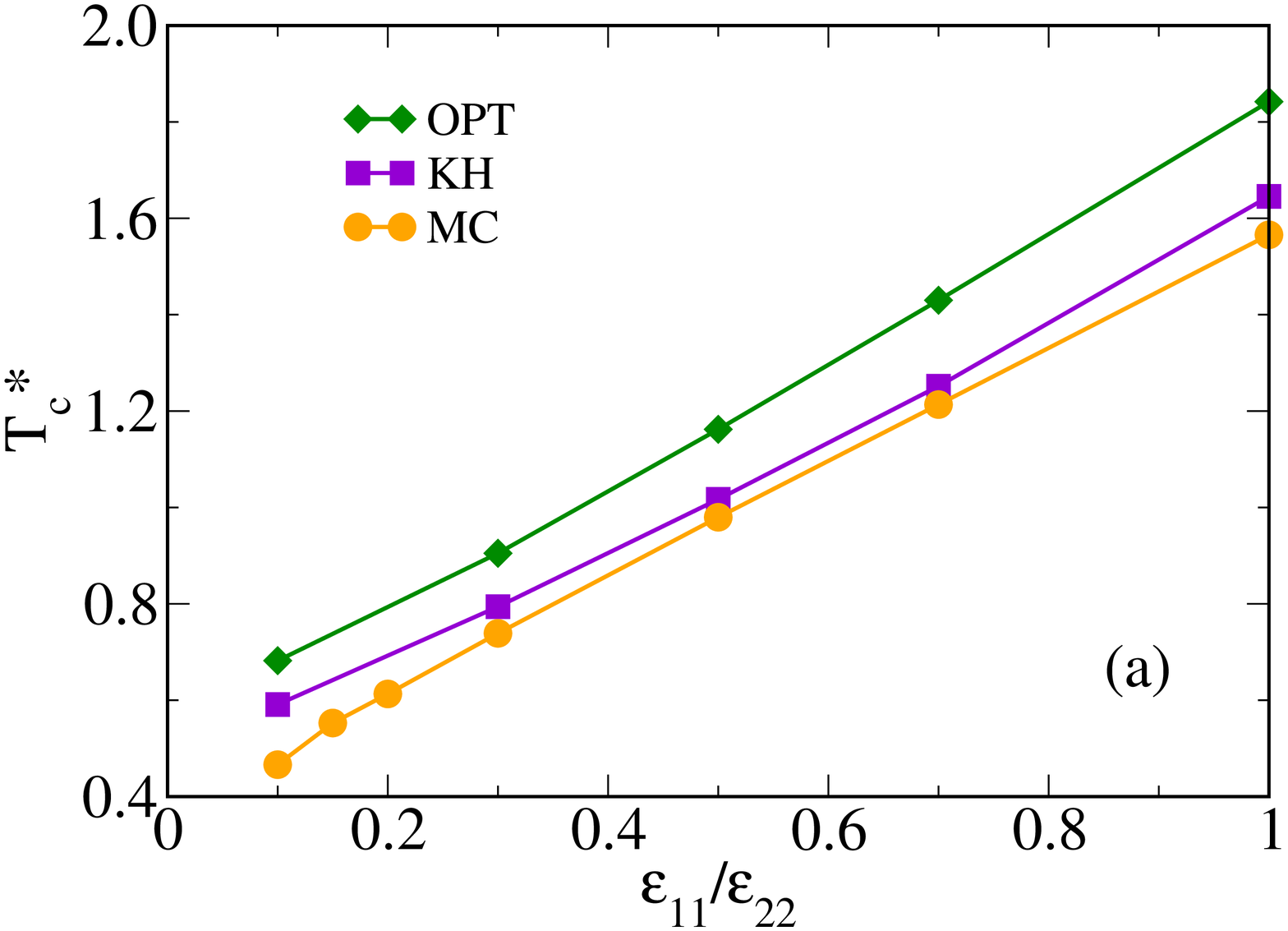}  \\
\includegraphics[width=9.0cm,angle=0]{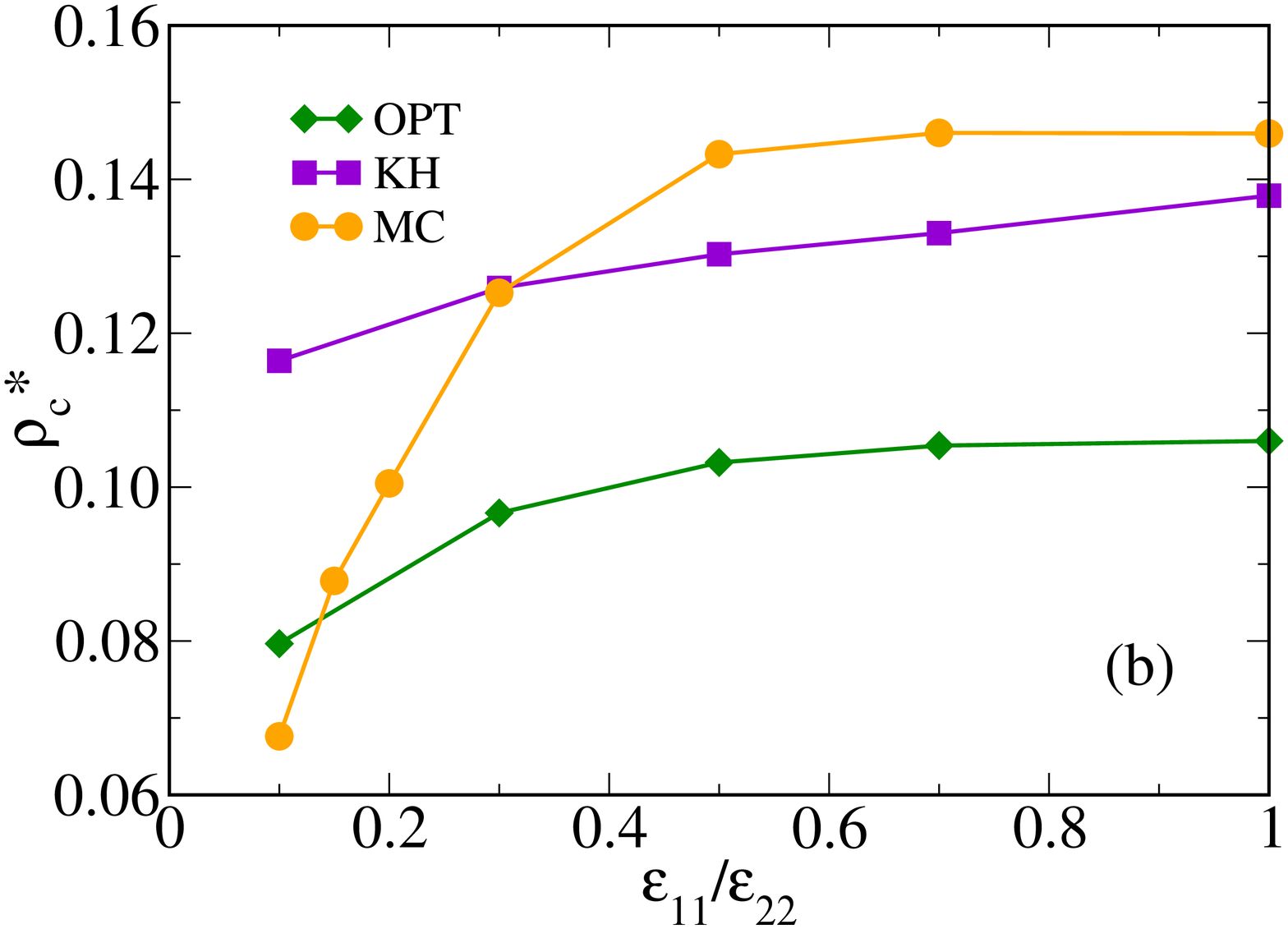} 
\end{tabular}
\caption{Critical temperatures (a) and densities (b) 
of colloidal dumbbells as functions
of $\varepsilon_{11}$.}\label{fig:Tc} 
\end{center}
\end{figure}

Numerical values of KH, OPT and MC 
critical parameters for all $\varepsilon_{11}$ investigated in this work
are collectively reported in Tab.~\ref{tab:crit}.
Critical temperatures and densities as functions of
$\varepsilon_{11}$ are shown in Fig.~\ref{fig:Tc}.
Remarkably, both theories agree with simulations in providing a linear
dependence of the critical temperature on 
$\varepsilon_{11}$. Such a 
circumstance is compatible with a mean field behavior,\cite{Munao:14} with
the only exception (for MC data) of the 
$\varepsilon_{11}=0.1\varepsilon_{22}$ case. 
Conversely,
the critical density $\rho^*_{\rm c}$ follows 
quite a different trend, keeping an
almost constant value from the symmetric case down to 
$\varepsilon_{11}=0.5\varepsilon_{22}$, 
with a marked drop 
upon further lowering this parameter.
Surprisingly enough, OPT qualitatively
reproduces such a behavior, 
actually predicting a critical density close
to simulation data for 
$\varepsilon_{11}=0.1\varepsilon_{22}$. 
On the other hand, the dependence of
KH $\rho^*_{\rm c}$ on 
$\varepsilon_{11}$ is less pronounced, 
failing to reproduce in particular the very low values 
attained at 
$\varepsilon_{11}=0.1\varepsilon_{22}$.

In a recent study,\cite{Munao:PCCP}
we have analyzed the RISM performances for
a close class of SW dumbbells,
characterized for instance
by an attraction range $\lambda=0.1$,
shorter than that investigated here (namely $\lambda=0.5$).
In that case we have shown that the HNC theory
is able to describe the low temperature regime
whereupon the gas-liquid coexistence takes place;
in particular, the theory predicts
the disappearance of a stable phase separation
for $\varepsilon_{11} < 0.4\varepsilon_{22}$, i.e. at
a value slightly larger than that observed in simulations 
($\varepsilon_{11} < 0.2\varepsilon_{22}$).
Also for $\lambda=0.1$
RISM agrees with simulations in predicting 
the linear scaling of the critical temperature discussed 
in Fig.~\ref{fig:Tc}a, even if with a larger discrepancy
with respect to the present case;
a better agreement is instead observed for the critical density,
essentially because this quantity, at $\lambda=0.1$,
keeps an almost constant value as a function of $\varepsilon_{11}$, 
at variance with the sudden drop documented in Fig.~\ref{fig:Tc}b.

\section{Conclusions}
We have presented a theoretical study of 
thermodynamic properties
and gas-liquid phase coexistence of anisotropic colloidal dumbbells,
where the anisotropy stems from the fact that,
although the two colloids forming the dumbbell
are identical in sizes, their interactions are not,
as one of the two attractive
square-well depth is gradually reduced to zero
toward the Janus dumbbells limit.
We have discussed and contrasted to each other 
the performances of
two venerable and well established tools,
the RISM and OPT theories of liquids.
As for the RISM, this theory
hinges on the
solution of the integral equations of liquids, 
thus providing the intermolecular
correlations, whereby all thermodynamic properties can be obtained.
Using the symmetric model as a test case, we have shown
that the Kovalenko-Hirata closure
(combining efficiently MSA
and HNC approximations) complemented with the energy 
route from structure to thermodynamics
is superior to the HNC approximation and other
procedures to calculate the thermodynamic properties.
As for the 
OPT theory, this method hinges on a fourth-order
high-temperature perturbative expansion of the free energy
of the square-well fluid, recently generalized to deal with molecular fluids.

The accuracy of both methods has been assessed 
by a direct comparison
with recent numerical simulations. 
We have found that both approaches provide reasonable 
predictions, 
with RISM slightly outperforming OPT
in almost
all cases with the notable exception of the critical density, 
where OPT is more accurate.
Latter predictions could be improved by finding a way to fully include
in the formalism the intramolecular correlations, presently taken into
account only at the level of the reference free energy.
 
We emphasize
the importance of having a reliable theoretical tool 
to provide a quick estimate
of thermodynamics of colloidal dumbbells. 
Recent experiments,\cite{kegel} as well as complementary numerical 
work,\cite{Munao:15}
show that in most of the interesting cases the two colloids forming 
the dumbbells are
asymmetric both in size and interaction. 
This means that a huge number of possible
combinations could be in principle studied to probe many 
topologically different
resulting phase diagrams. 
A reliable theoretical approach would then be invaluable in
pinning down the most interesting cases where a more 
sophisticated, and computationally more demanding,
calculation could be performed.

\acknowledgments
GM, DC, CC, FS, and AG gratefully acknowledge support from 
PRIN-MIUR 2010-2011 Project.


%

\end{document}